# Quantum sensing with atomic, molecular, and optical platforms for fundamental physics*

Jun Ye

JILA, National Institute of Standards and Technology, and Department of Physics, University of Colorado, Boulder, Colorado, 80309, USA

Peter Zoller

Institute for Theoretical Physics, University of Innsbruck, 6020 Innsbruck, Austria
and Institute for Quantum Optics and Quantum Information, Austrian Academy of Sciences, 6020 Innsbruck, Austria

Atomic, molecular, and optical (AMO) physics has been at the forefront of the development of quantum science while laying the foundation for modern technology. With the growing capabilities of quantum control of many atoms for engineered many-body states and quantum entanglement, a key question emerges: what critical impact will the second quantum revolution with ubiquitous applications of entanglement bring to bear on fundamental physics?

In this Essay, we argue that a compelling long-term vision for fundamental physics and novel applications is to harness the rapid development of quantum information science to define and advance the frontiers of measurement physics, with strong potential for fundamental discoveries.

As quantum technologies, such as fault-tolerant quantum computing and entangled quantum sensor networks, become much more advanced than today's realization, we wonder what doors of basic science can these tools unlock? We anticipate that some of the most intriguing and challenging problems, such as quantum aspects of gravity, fundamental symmetries, or new physics beyond the minimal standard model, will be tackled at the emerging quantum measurement frontier.

*Part of a series of Essays which concisely present author visions for the future of their field.*

**Introduction.**— The study of individual atoms, real or artificial, has brought us a tremendous understanding of fundamental physics such as the vacuum mode structure of electromagnetic fields and has given birth to powerful tools such as lasers and nuclear magnetic resonance (NMR) [1-5]. The tantalizing possibility of assembling and controlling thousands to millions of atoms as precisely as a single atom, bringing them together for engineered many-body states and quantum entanglement, is guaranteed to provide technological breakthroughs in information collection, processing, and communication (Fig. 1). While quantum computing is still in a very early stage of development, the implication on how quantum systems collect, process, and distribute information (for sensing and communication) is already fundamentally shaping the scope and impact of AMO science well beyond its traditional boundary [1].

AMO platforms feature both spin qubits and qudits, and bosonic and fermionic particles as basic constituents [6], and their functionality can be tailored to specific measurement goals [7]. Atoms and molecules can now be individually addressed and measured with high fidelity [8-10], then interconnected into small systems decoupled from the environment [11, 12]. The development of increasingly large-sized and programmable quantum systems with a mesoscopic number of particles will require both creative scientific ideas and innovative technology development [Fig. 1(a)]. Equally important is the identification of near-term applications based on these machines, and together these developments will fuel the exploration of new physics beyond the current reach. For example, atomic interferometers built to sense either the internal (clock) [13, 14] or external (inertial) [15, 16] degrees of freedom [Fig. 1(b)] can employ entanglement to reach beyond the standard quantum limit defined by uncorrelated atoms [17-19], and the resulting Heisenberg limit will allow precision to scale directly with the number of atoms, an outcome representing the best performance permitted by quantum physics.

Chains of trapped ions were among the first systems to realize digital quantum computing that features high gate fidelity [20]. With enhanced connectivity they have recently been used to run quantum circuits implementing quantum algorithms [21], as well as digital and analog quantum simulations of complex Hamiltonians [Fig. 1(c)] [22]. Neutral atoms are rapidly rising to realize medium-scale quantum processors, including scaling to tens of thousands of atoms, with local addressing and atom storage and reloading [Fig. 1(d)]. The increasingly higher fidelity and programmability for one- and two-qubit gates, along with effective midcircuit readout, and flexibility in spatial reconfiguration and entanglement distribution in a network of qubits, are leading to the production of encoded and error-corrected logical qubits, permitting the systems to make fault-tolerant production of logical GHZ states that can be used to measure certain parameters at an enhanced rate [23].

Indeed, quantum simulation and quantum sensing are inextricably linked [24]. Simulating many-body quantum systems provides fundamental insights into dynamical phases of matter, and this knowledge can be leveraged to design a new generation of quantum sensors with engineered many-body states and collective quantum measurements [25-27]. The goal is to develop sensors that supersede the current state of the art for targeted applications that are specifically motivated by the need to probe fundamental physics [28]. It is through such applications that current intermediate-scale quantum technologies can have a great impact on both fundamental science and disruptive technology.

To realize a meaningful quantum advantage [29], quantum technologies must demonstrate scalability and superior performance in regimes where classical approaches reach their limit, and it is necessary to characterize and verify quantum devices in the classically inaccessible regime [30]. Demonstrations of quantum advantage on problems that have practical applications, such as entanglement enhanced quantum sensors, will help expand the deployment of quantum devices in the field, which will in turn open new scientific opportunities. Achieving that requires demonstrations of robustness and compactness for field applications including the distribution of entanglement in a network [16, 31, 32]. It is further important to improve our ability to perform and characterize measurement in multiple-parameter quantum devices, such as sensor networks [33, 34]. The close coordination of quantum hardware with quantum algorithms and measurement protocols to optimally extract relevant physical information [35], together with identification of impactful applications that are best suited for these quantum devices, will be critical for demonstrating a quantum advantage for meaningful applications.

*Emerging quantum technologies based on AMO platforms*.— Deploying entanglement will significantly increase the measurement sensitivity at a given bandwidth, or equivalently, the sensing bandwidth for a given precision. One can weigh the desired properties of a quantum sensor against the available quantum resources provided by a specific platform and design relative quantum circuits to enable entanglement and collective measurement capability. For example, parallel production of various sizes of maximally entangled GHZ states can help realize Heisenberg limited metrology [Fig. 1(e)] [36], and variational optimization algorithms [37, 38] can be used to approximate these quantum circuits to achieve robust performance by providing simultaneous enhancement in bandwidth, sensitivity, and dynamic range of a desired physical quantity to be measured [39]. This approach may be particularly beneficial for future searches involving weak signals at undetermined frequencies such as those of astrophysical origins like dark matter [40] and gravitational waves [41, 42].

Improving fidelity for atom-light interactions at the single quanta level also represents a critical piece of technology needed to build long-distance entanglement for a distributed quantum sensing network [Fig. 1(a)] [31, 32, 34]. We can devise new and powerful ways to measure and control electromagnetic fields based on the optimization of atomic quantum resources [43], and this will in turn further improve our capabilities for quantum system manipulations. For example, we can connect a large number of miniature optical cavities around individual atoms or a collection of atoms to allow coherent exchange of superposition and entanglement while suppressing fundamental decoherence [44, 45]. We can confine individual atoms in a specially configured geometry to achieve a phased array of atomic antennas that will allow a many-body quantum state to modify the vacuum mode structure surrounding atomic qubits, thus enabling the generation, propagation, and detection of photons that are engineered to achieve high-fidelity atom-photon interaction. Exotic quantum states with specific metrological purposes can be designed based on this approach, including subradiant states [46] that could enhance quantum coherence unless perturbed by a desired physical signal we wish to monitor. The realization of high-efficiency control of quantum states of light will help overcome a key bottleneck of the current shortfall of quantum networks due to the lack of high-fidelity quantum repeaters and/or quantum transduction devices.

It is also highly desirable in many applications to extend the phase coherence of electromagnetic waves far exceeding the current limits. This can be realized using a cascaded array of entangled atoms with varying sensitivity and dynamic range to the interacting field. With high-fidelity entangled states realized for successive groups of a growing number of atoms, the progressively increased parity oscillation frequencies can be exploited to guide the phase coherence of an electromagnetic oscillator over a large dynamic range [Figs. 1(d), 1(e)].

The discussion of emerging quantum technology would be incomplete without raising interesting questions on fundamental physics [47]. As we list a few examples that are worthy of exploration in our opinion, we note the increasing importance of collaboration between theorists and experimentalists. The field of AMO and quantum science is now intricately intertwined with condensed matter, high-energy, and chemical physics as well as astrophysics and information science. This provides both tremendous challenges and outstanding opportunities to identify deeply insightful and broadly applicable connections with clear goals for making important, potentially groundbreaking, discoveries. Quantum technologists need to work with particle physics theorists to explore powerful searches for new physics, with condensed matter theorists on the emergent phases of matter and their implications for metrology, with algorithm engineers to design and implement optimal measurement protocols, and with molecular scientists on a broad scope of important problems concerning chemistry and biology.

***Connections and application to fundamental symmetry and gravity.*** —

***Atomic clocks.*** — There are a few natural questions to ask: How far can we continue to advance the performance of today's atomic clock [2]? What is the ultimate limit of an atomic clock? Will the clock performance scale with what is maximally allowed by quantum mechanics [36, 48]? Can we combine high-precision interferometry for both internal degrees of freedom (clock) and external degrees of freedom (matter-wave interferometer) to directly probe connections between quantum physics and gravity [49, 50]? Matter-wave interferometers are proposed to search for gravitational waves at intermediate frequencies [51-53]; can clocks play an important role in the future gravitational-wave observatory at a unique frequency band as well [41, 54]? Can clocks detect space-time curvature and confront the conflict between determinism and quantum decoherence [55]? Will a distributed network of clocks, synchronized via classical or quantum means, aid the search for the elusive dark matter in a critical way [56, 57] [Fig. 1(a)]? Studying the quantum nature of clocks may also shed insights into topics of thermalization, dynamics, and work in an open quantum system [58].

Regarding the question of what the ultimate limit for metrology of time is, we note that recent advances in optical atomic clocks have demonstrated measurement precision to better than 1 part in $10^{20}$ [59], along with accuracy of better than 1 part in $10^{18}$ [60, 61] [Fig. 1(b)]. Looking at the quantum engineering of the atomic clock system, a number of important ingredients emerge that would enable continued advances of clock performance in precision without hitting a fundamental limit for the next few decimal points. A continuous flow of quantum-state-prepared atoms into a clock platform will allow uninterrupted frequency measurement operation along with a continuously phase-steered optical local oscillator. An increasingly large number of atoms will enhance the clock precision in principle, but we will need to understand the underlying many-body Hamiltonian and explore its impact on quantum coherence and frequency shift. Relatively weak interactions will start to manifest themselves as important systematic effects that may potentially limit clock performance until they are understood microscopically. Here we have a strong intellectual overlap between quantum simulation of complex phases of matter and quantum metrology, with the final goal to build powerful quantum platforms to discover, understand, and utilize many-body Hamiltonians that are metrologically powerful and lay the foundation for a new generation of optical atomic clocks. This process will provide critical insights and give rise to new measurement protocols to ensure the number of atoms and the measurement precision advance side by side. Entanglement generation and employment will be integrated into the natural evolution of the clock measurement sequence, with a clear understanding and control of possible consequences on measurement errors.

Overcoming known technical limits benefits largely from the measurement speed, so precision helps in a big way. With the clock being a measurement device for space and time, the remaining issues would naturally be associated with gravity [62]. As usual, when a particular effect becomes dominant in the performance of a quantum device, it turns into a sensitive sensor for such an effect. Considering that the gravitational redshift is $1 \times 10^{-19}$ for a 1 mm height change, when we improve the clock performance to $1 \times 10^{-21}$, we will be able to measure gravitational effects on clocks with atoms separated by a mere 10 μm. With spatial superposition extending to this length scale, we will measure the gravity effect on quantum coherence and entanglement, a feat never achieved before. The performance of $1 \times 10^{-21}$ for clocks is also a major milestone that signifies the possibility of employing atomic clocks directly for gravitational-wave detection at a frequency range that is complementary to LIGO. Overcoming challenges such as integrating atomic system with photonics, we can take clock technology out of the laboratory to play an important role in future relativistic geodesy research, where the next generation of optical atomic clocks will help characterize and define the time-dependent properties of the Earth's gravity field [63, 64].

Thinking of the problem more generally, we wonder whether the future clocks will eventually turn into a precise probe for the interface of quantum coherence and gravitational effect [65], especially with the technology of entangling clocks distributed across a significant distance along the direction of gravity [Fig. 1(b)]. This is a very interesting area to investigate. When the clock measurement precision reaches the level of $10^{-22}$, its frequency becomes sensitive to a spatially extended quantum superposition if it spans 1 μm along the direction of gravity. Even the momentum spread of various motional eigenstates of atoms confined in optical lattices will have measurable motional time dilation effects on quantum state evolutions. Therefore, as we drive the clock transition to create superposition in both internal and external states, we will need a relativistic Hamiltonian that explicitly incorporates spin-dependent atomic mass and position-dependent gravitational phase, as well as motional eigenstates. Fundamentally new dynamics will emerge when we can observe them with enhanced clock precision.

The relation of gravity and many-body physics in the lattice is intriguing. As we drive an optical clock transition from ground state $|g\rangle$ to excited state $|e\rangle$, the atomic mass $M$ must be corrected by the mass defect with respect to the internal binding energy, i.e., $(M+E_e/c^2)|e\rangle\langle e| + M|g\rangle\langle g|$. A gravitational potential $V_{grav}= M g z$ leads to a coupling of the internal states and the spatial wave function. The corresponding relativistic Hamiltonian contains corrections of both motional and gravitational time dilation effects to the transition frequency. The higher the transition frequency $E_e = h\nu$, the larger the motional time dilation effect. For an optical clock transition such as that in $^{87}$Sr with atoms cooled to the motional ground state inside an optical lattice, this motional time dilation effect can be on the order of 4 x $10^{-22}$. Meanwhile, the gravitational redshift between each lattice site along gravity (~407 nm using the Sr magic wavelength for trapping) is about 4.4 x $10^{-23}$. Hence, when we prepare a coherent superposition or entanglement of spin and motional wave functions across a spatial separation of several micrometers, involving two or more motional bands of the lattice, we will observe new phenomena on the clock evolution arising from genuine general relativity effects on both spin and motion. Now imagine a gravitation-mediated spin-orbit coupling in a quantum many-body system confined in a lattice spread across a gravitation potential. We will be entering unexplored territory for physics, if we have sufficient clock precision to observe the related dynamics. Pushing the spatial extent of a distributed clock network will allow us to explore space-time curvature beyond linear gravity, and larger energy separations between clock states (for example, nuclear clock instead of the current optical clock) allow a more stringent test of motion and decoherence arising from general relativity.

Can we extend the precision and capabilities of the current generation of optical lattice clocks to perform these new tests of the foundations of relativity and quantum physics at unprecedented sensitivity? We should be optimistic as we continue to advance the state-of-the-art control of both internal and external coherence over macroscopic timescales and spatial separations. It is tantalizing to put thought experiments based on relativistic principles into a quantum reality so that we can explore both the gravitational and motional time-dilation-induced decoherence when we have joint superposition or entanglement in the internal spin and external motional degrees of freedom.

Interactions between many atoms in these systems will also provide new dynamics to enhance the detection of general relativity effects on quantum coherence and entanglement [66].

*Permanent electric dipole moment.*—The search for this quantity represents a profound test of fundamental symmetry of time reversal (*T*), which is equivalent to charge-parity (*CP*) by the *CPT* theorem. The completion of the standard model of particle physics with the discovery of the Higgs boson only thickens the plot that there is new physics beyond the standard model. Many mysteries abound in our Universe that remain unaccounted for by the standard model: the microscopic origin and nature of dark matter, the matter-antimatter asymmetry, and the hierarchy problem, among others [7, 67]. Discovery of new particles at the mass scale >10 TeV $c^{-2}$, beyond the direct reach of the LHC, could account for the observed but unexplained phenomena. These new particles can provide time-reversal symmetry violation that is much stronger than what has been observed so far. A permanent electric dipole moment (EDM) of nucleons and electrons is a source of such *T* violation, and thus searches for EDM through low-energy precision measurements of atoms and molecules have gained much interest and progress [68]. The most recent upper bound for an electron's electric dipole moment in such an experiment is placed at $4.1 \times 10^{-30}$ *e* cm, which can be interpreted as reaching beyond 10 TeV $c^{-2}$ mass scale [69].

One may naturally ask whether and how the current quantum revolution is going to benefit this most fundamental probe of our Universe. Can we imagine quantum technology will bring future experimental limits on *T*-violating beyond-standard-model physics to the 1000 TeV eV scale? Recent progress in AMO-based searches for the electron EDM has built on *quantum state control of molecules*, a new emerging field inspired by the success of quantum science with ultracold atoms [70]. If quantum technology on atoms is giving rise to a dramatic enhancement in measurement precision and a widened list of applications to fundamental physics such as searches for dark matter, tests of general relativity, and symmetry such as parity violation, then we have all the reason to be optimistic that a parallel revolution will take place in molecular physics where the unique properties of molecules designed for the specific purpose of EDM searches will be put to the most optimal use via quantum technology. Along the way, many outstanding challenges and opportunities will arise, and the impact of these new technologies will only widen to include other areas of physical sciences such as chemistry.

The stringent requirements of high-energy resolution and strong relativistic effects for EDM-sensitive systems have stirred up our imaginations for an increasingly long list of suitable atomic and molecular candidates and the corresponding creative approaches for using them to reach unprecedented precision. Generally, quantum technology will provide some of the most important ingredients: suitable atomic species or designer molecules that feature a greatly enhanced EDM sensitivity, quantum state manipulation of the target atoms or molecules for the longest possible coherence time, employment of a large number of such atoms and molecules in a single measurement or in continuous measurement whenever possible, and finally, application of spin entanglement to increase the detection bandwidth and to sidestep technical limitations on coherence. The diversity of search candidates and the variety of search approaches are invaluable for placing the most stringent constraints or for confirming a non-null result, along with the prospect of cross-fertilization and further advancing emerging quantum technologies for a broader set of physical platforms, some of which may be critical for improving global constraints on beyond-standard-model physics by looking for a *CP*-violating nuclear magnetic quadrupole moment [71].

An exciting recent development is the laser cooling of candidate molecules that are specifically designed for high EDM sensitivity while remaining relatively easy for light-based quantum state manipulation and featuring systematic effect-rejecting energy level structure [72]. Even polyatomic molecules have been brought to ultralow temperatures that are compatible with optical trapping [73]. In the long run, technologies such as state-independent trapping for electron spin resonance spectroscopy, a continuous beam of state-prepared molecules, suppression of molecular interaction and loss, spin squeezing, and high quantum efficiency state readout will be developed, and the sensitivity for the EDM search can be dramatically increased by orders of magnitude from today's best, unveiling a very exciting epoch of doing particle physics research at the ultralow energy regime, where quantum physics plays a dominant role.

*Dark matter*.—The advances in atomic clocks and EDM searches can also contribute to the search for dark matter of ultralight mass [74, 75]. Candidates for ultralight dark matter particles include the axion, motivated by QCD, and the dilaton, which is predicted by string theory as a scalar partner of the tensor Einstein graviton. Light scalar fields from extra dimensions offer a solution for the hierarchy problem. Astrophysical observations constrain the ultralight dark matter to be in the form of a bosonic field with the Compton frequency proportional to the mass. The axion(-like) dark matter may couple to the electron, leading to an oscillatory signature in the electron EDM. The dilaton [76] could also be detected via possible scalar couplings to the four fundamental forces of the standard model, which cause oscillatory perturbations to the effective masses and gauge couplings of fundamental particles. In particular, the electromagnetic coupling term could cause oscillations in the fine-structure constant $\alpha$. Additionally, the gluonic, quark, and electron mass coupling terms could also cause variations in the proton-to-electron mass ratio [77].

Comparisons of clocks based on different atomic species with varying levels of sensitivity to a potential time-dependent variation of $\alpha$ lead to an effective search for the oscillatory signature of an ultralight dark matter field [40, 56, 78]. This signature will be a narrow-linewidth signal in frequency space, with a maximum width of $10^{-6}$ of the dark matter mass with the coherence given by the virial velocity from astrophysical observations. For typical observation sampling rates and data collection durations, we can search for ultralight dark matter oscillations at the 1 µHz to 1 kHz scale, corresponding to a mass range of $4\times10^{-21}$ to $4\times10^{-12}$ eV/$c^2$. This represents a large search space, and quantum technologies become highly desirable. For example, the use of multiple pulses during a coherent spectroscopy sequence allows a researcher to tune the sensitivity function across a desired frequency range. The use of spin entanglement can be particularly effective in improving the measurement precision at finite bandwidth. There is also great room for sensitivity enhancement on searches by taking advantage of state-of-the-art optical clocks and ultrastable reference cavities [79]. A direct frequency comparison of an optical clock against a laser referenced to a highly stable optical reference cavity is linearly sensitive to variations in $\alpha$, and this search precision is anticipated to improve rapidly in the future. Finally, when the clock network is established with state-of-the-art performance [57], we will have in hand an effective large-scale interferometer with which we can search for spatial and time-dependent dark matter signals.

***Connection and applications to condensed matter, biological, and chemical physics.*** —

*Condensed matter physics.*—Quantum sensors, especially those based on entangled atoms or molecules, are quantum systems designed to achieve maximum sensitivity to external parameters and signals of interest that act as external disturbances on these systems. This is intrinsically linked and synergistic with problems in condensed matter (CM) theory and experiment, which deal with strongly interacting many-body systems involving a large number of particles [80]. CM focuses on phases and phase transitions as equilibrium phenomena, or out-of-equilibrium quench dynamics involving thermalization and scrambling, and the generation of highly entangled quantum states. CM's primary goal is understanding and designing quantum materials. In contrast, AMO physics creates and employs such many-body systems as quantum simulators, or ultracold atomic and molecular quantum matter with engineered interactions, and investigates them using the unique tools of quantum optics [1, 24].

The central role of spins, real or pseudo-, in quantum sensing naturally links it to quantum metrology through the exploration of many-body spin physics, covering topics such as spin liquids, quantum magnetism, and more broadly, superconductivity. The paradigmatic CM model that describes phenomena such as quantum magnetism is the Heisenberg model. Such interacting spin models that can be utilized as a resource for generating metrologically useful entanglement are natively implemented in AMO quantum simulators such as trapped ions, Rydberg atoms in tweezer arrays, or as Bose and Fermi Hubbard models with atoms or molecules in optical lattices [1, 24]. For example, Heisenberg models emerge as effective spin models from a range of interactions that include contact, superexchange, and dipolar as featured in present optical lattice clock setups in various geometries.

The dynamics of quenching with long-range spin-spin interactions lead to the creation of spin-squeezed states that have meteorological relevance. The quintessential example of this is provided by infinite-range spin-spin interactions in one-axis and two-axis twisting scenarios. These infinite-range interactions are naturally manifested as Molmer-Sorensen interactions [81] in trapped ion quantum computing and in cavity QED setups. However, analog quantum simulators typically involve finite-range interactions, such as dipolar interactions with polar molecules, van der Waals interactions between laser-excited Rydberg atoms, or trapped ion quantum simulators with tunable power-law spin-spin interactions [17-19, 24]. The challenge lies in engineering nonequilibrium dynamics that involve Floquet driving or Trotterized time evolution from the available entangling interactions to achieve spin squeezing in large-scale quantum systems. As elaborated below in the discussion of theoretical connections between quantum information and quantum metrology, these finite-range spin-spin interactions also provide a natural resource to engineer quantum circuits for variational algorithms in quantum metrology.

Equilibrium quantum phases in *XY* models of quantum magnetism can provide a natural resource for spin squeezing with metrological relevance [82]. Quantum phase transitions sensitive to external physical parameters could be the key to unlocking new possibilities in the field of quantum sensing, providing a pathway for the development of more precise and robust sensors. Quantum phase transitions, which are transitions between different quantum phases at zero temperature by tuning of an external control parameter [80], feature long-range entanglement, symmetry breaking, and gap closing and exhibit quantum enhancement for sensing at criticality [83]. Interestingly, quantum enhanced sensing is possible using topological edge states near the phase boundary, even in systems that exhibit neither symmetry-breaking nor long-range entanglement, providing the opportunity for topological quantum sensors to promise robustness against local perturbations [84].

***Biological and chemical physics.***—The intimate connection between quantum metrology and condensed matter naturally expands to a broader scope of material science that overlaps with chemical and biological physics [85]. Quantum control of atom-like impurities inside a solid lattice provides a remarkable bridge between atomic and CM qubits, turning these artificial atoms in solids into unique quantum sensors. With a judicious balance of isolation and strong interactions, we can manipulate the quantum coherence of these solid defects even under ambient conditions and engineer their couplings to photons and phonons or their surrounding spin bath.  Examples of artificial atoms in solids include nitrogen-vacancy (N-*V*) or silicon-vacancy (Si-*V*) centers in diamond, defects in silicon carbide, and two-dimensional materials. N-*V* or Si-*V* centers feature robust quantum coherence and have been used to realize high-quality quantum memory and multiqubit quantum registers.  The rapid development of this CM platform for quantum science and metrology is finding practical applications in biomedical diagnostics through, for example, the development of highly sensitive, miniature magnetometers [86] to record and analyze NMR signals from individual molecules and single cells [87].

Molecules offer unique advantages for certain sensing applications, such as the search for EDM or dark matter, and they will play an increasingly important role in fundamental physics as the related quantum control technology gets developed and matures. Hamiltonian engineering for quantum sensing, for example, with spin squeezing via single-axis or two-axis twisting, can be implemented for molecules just as optimal quantum control pulse sequences have been developed for atoms and N-*V* centers [88].  At the same time, the capabilities of manipulating individual molecules and a group of molecules, all with precisely engineered quantum states, offer the ultimate assembly line for carefully choreographed, step-by-step controlled chemical reactions. Indeed, the theme of chemical physics is to understand and control how molecules transform from reactants into products. The idea of completely controlling and probing chemical reactions would require the reactant molecules to be prepared in individual quantum states, the collisional process maintained in quantum channels guiding how molecules approach each other [89], the transient complex formed during the reaction fully monitored and even controlled [90], and the product molecules characterized in all degrees of freedom. The initial state preparation and readout are increasingly precise and sophisticated with the development of cold molecules [70]; however, the large range of energy scale and the associated complexity throughout reaction coordinates pose significant challenges to the existing quantum technology.

Applying quantum metrology tools to molecules will help us find ways to observe and manipulate intermediate products during the reaction, assisted with the identification of resonant or transition states of the collision complex. Observing, characterizing, and controlling these resonances will shed light on how atoms share and transport energy, sometimes assisted with external control fields, providing insights into the chemical process. Quantum metrology will also assist reaction product investigation, which is to ascertain which chemical species emerge, along with their internal states and external motional degrees of freedom. These tasks would require spectroscopic accuracy in measuring both the transition and product states, coupled with necessary time resolution to determine their respective populations in real time.  In recent years such a tool, with broad spectroscopic capabilities and exquisite time resolution, has indeed been developed in the form of infrared frequency comb spectroscopy [91]. Watching the population of the relevant states as a function of time has already led to powerful applications in medical science, for example, in breath-based diagnosis of medical conditions [92].

### *Theory connections between quantum information and quantum metrology*. —

The goal of quantum metrology is to identify and build the most precise quantum sensors permissible within quantum physics, leveraging entanglement as a resource to exceed the standard quantum limit for uncorrelated particles. Traditionally, the notion of "optimal" quantum sensors revolves around saturating fundamental precision bounds set by quantum parameter estimation, e.g., as a Cramer-Rao bound, where the (quantum) Fisher information defines the utmost achievable precision for a given parameter value [93]. Taking a broader perspective, we can conceive the optimal quantum sensor to be defined via optimization of a metrological cost function tailored to a specific measurement task, e.g., as an optimal signal to noise ratio over a given parameter range as exemplified in atomic clocks and magnetometry [39]. This applies to single- and multiparameter estimation and might consider regimes involving single, few, or many-measurement scenarios of parameter estimation [93-95]. Adopting the language of quantum information, we can phrase this as a variational algorithm for quantum metrology [37, 38, 96, 97], where we determine optimal entangling and decoding quantum circuits defining probe states and measurements. A practical approach towards the optimal quantum sensor might thus consider constrained, finite-depth quantum circuits feasible on a given sensing platform [39, 98], as facilitated, for instance, in a scalable manner by programmable quantum simulators with interaction between particles as entanglement resource [24].

This quest for the optimal variational circuits can be executed "on the quantum device" in a quantum feedback loop, thus finding the optimal operating point in the presence of device decoherence, while capitalizing on the robustness of variational algorithms to coherent control errors [39, 98]. We note that optimization on the quantum sensor solves a classically hard problem involving many-body quantum dynamics. Hence, optimization on the programmable quantum sensor, which first identifies the optimal operational point and consequently utilizes the optimal quantum circuits for metrology, serves as an illustration of achieving a practical quantum advantage [29] in a quantum metrological application. We expect this concept of defining and approximating optimal sensors variationally is rather generic, and extends to discovering and distributing optimal entanglement to measure global field properties in quantum sensing networks [32, 33, 99].

At the confluence of quantum information and quantum metrology, we expect quantum error correction, as one of the most astounding achievements in theoretical and experimental quantum information, to have the potential of overcoming and mitigating errors in quantum metrological protocols [100]. Another example is Hamiltonian learning in quantum simulations [30, 101-103], which we can view as a quantum metrological task, estimating simultaneously the—*a priori* unknown—coupling parameters of a many-body Hamiltonian from measurements on the quantum device, in a sample-efficient way and scaling. Finally, beyond estimation of classical parameters, recent progress in classical shadows and randomized measurements illustrates sample-efficient learning of quantum states and many-body observables, opening new avenues for quantum metrology [104].

***Concluding remarks.—***Today's excitement for quantum science brings along a new set of opportunities and challenges: engineer interactions and correlation between many quantum objects, whose wave functions may extend over macroscopic spatial and temporal intervals and employ quantum coherence, superposition, and entanglement for specific measurement goals. Precise control of distributed quantum systems represents both a fundamental challenge and an emerging resource for a novel class of quantum-enabled sensors that could give rise to groundbreaking scientific opportunities and would impact all areas of physical sciences. In this Essay, we have touched on only a small set of examples of the exciting prospect of quantum metrology for fundamental and applied physics. In the future, as different scientific goals will certainly emerge and new quantum platforms and measurement protocols will arise [83, 105, 106], a common theme will connect them by maximizing the recovery of quantum Fisher information [35] through optimal design and efficient use of quantum resources to achieve the best possible advantage of the desired scientific signal against fundamental quantum noise.

***Acknowledgments.***— We thank Steven Bass, A. Gorshkov, R. Kaubruegger, and S. Lannig for providing feedback on the manuscript. We are also grateful to the help on figure artwork from S. Burrows, R. Kaubruegger, C. Roos, and D. Vasilyev. Finally, we acknowledge support from NSF Grant No. QLCI OMA-2016244 and Quantum Systems Accelerator of DOE Office of Science.

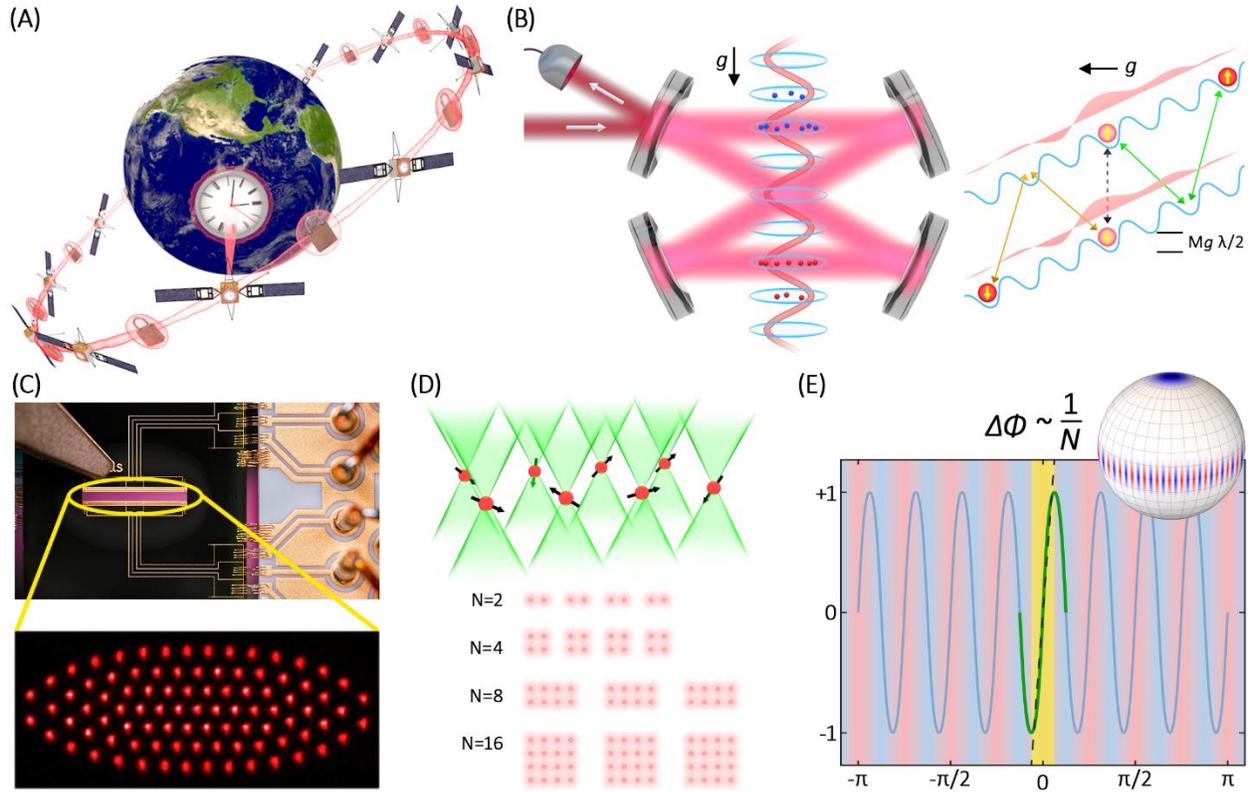

FIG. 1. Quantum system design and engineering to open new frontiers for quantum sensing. (a) A quantum network that links individual clocks in space with enhanced performance and security based on their entangled quantum states. The optimal use of the global resources and quantum-enabled precision and accuracy also represents a unique long-baseline observatory for fundamental physics [57]. (b) A Wannier-Stark optical lattice where clock (spin) and atom interferometer (motion) are integrated into a single quantum platform [59]. Cavity-QED-based entanglement generation will further enhance the probing of clock frequency, coherence, and gravity [14]. (c) A new monolithic ion trap configuration allows a two-dimensional arrangement of laser-cooled ions for quantum simulations of spin models and increased sensing capabilities [107]. (d) A tweezer array of neutral atoms that enable any-to-any connectivity among hundreds of atomic qubits with universal local single-qubit rotations and high-fidelity two-qubit Rydberg gates. Fast midcircuit readout and feedforward can be implemented together with parallel transport in reconfigurable array architecture [23, 108]. (e) The generation of GHZ entangled states for enhanced interferometric sensitivity in clock operation, reaching the Heisenberg limit where the phase sensitivity scales with the inverse of particle number $N$. This is the best possible outcome defined by quantum physics [36, 108].

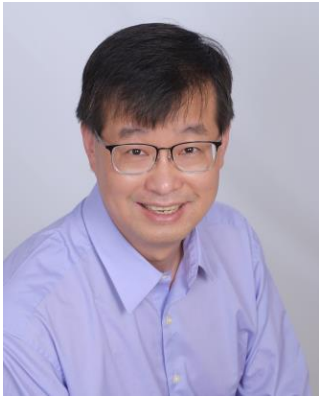

Jun Ye is a JILA and NIST Fellow and a member of the National Academy of Sciences. His research focuses on new tools for light-matter interactions and their applications for precision measurement, quantum science, and frequency metrology. He is known for developing highly precise and accurate atomic clocks, for the first realization of a quantum gas of polar molecules, and for pioneering work on frequency combs and spectroscopy. Professor Ye has received numerous honors including five Gold Medals from the US Department of Commerce, the Breakthrough Prize in Fundamental Physics, the Micius Prize, the Herbert Walther Award, the Vannevar Bush Fellowship, the Norman F. Ramsey Prize, and the I.I. Rabi Prize.

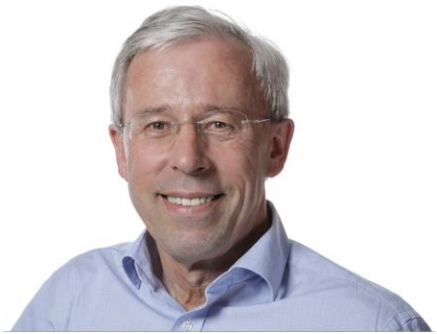

Peter Zoller is Professor *em.* at the University of Innsbruck and group leader at the Institute of Quantum Optics and Quantum Information of the Austria Academy of Sciences. He works on quantum optics and quantum information. He is best known for his pioneering research on quantum computing and quantum communication, for bridging quantum optics and condensed matter physics, and for his theoretical works on the interaction of laser light and atoms. Professor Zoller has received numerous awards for his contributions to physics including the Wolf Prize, the Dirac Medal, the Benjamin Franklin Medal, the Max Planck Medal, and the Micius Prize.